\documentclass[fdp,a4paper,fleqn%
]{w-art}
\usepackage{times,cite,w-thm}
\theoremstyle{plain}

\theoremstyle{definition}

\usepackage[]{graphicx}

\usepackage{amsfonts,amssymb, amsmath}
\usepackage[top = 2.0cm, bottom = 2.0cm, left = 2.0cm, right = 2.0cm]{geometry}

\newcommand{\extd}{\mathrm{d}}

\newcommand{\so}{\mathfrak{so}}
\newcommand{\su}{\mathfrak{su}}

\newcommand{\SO}{\mathrm{SO}}
\newcommand{\SU}{\mathrm{SU}}

\begin{document}
\DOIsuffix{theDOIsuffix}
\Volume{55}
\Month{01}
\Year{2007}
\pagespan{1}{}
\Receiveddate{XXXX}
\Reviseddate{XXXX}
\Accepteddate{XXXX}
\Dateposted{XXXX}
\keywords{List, of, comma, separated, keywords.}



\title[Quantum geometry in GFTs]{Non-commutative quantum geometric data in group field theories}


\author[D. Oriti]{Daniele Oriti\inst{1,}%
  \footnote{E-mail:~\textsf{daniele.oriti@aei.mpg.de}}}
\address[\inst{1}]{MPI f\"ur Gravitationsphysik, Albert Einstein Institut, Am M\"uhlenberg 1, D-14476 Potsdam, Germany}
\begin{abstract}
We review briefly the motivations for introducing additional group-theoretic data in tensor models, leading to the richer framework of group field theories, themselves a field theory formulation of loop quantum gravity. We discuss how these data give to the GFT amplitudes the structure of lattice gauge theories and simplicial gravity path integrals, and make their quantum geometry manifest. We focus in particular on the non-commutative flux/algebra representation of these models.
\end{abstract}
\maketitle                   





\section{Introduction} 
The idea of a fully background independent formulation of the microscopic structure of quantum space-time has recently found a tentative realisation in the (tensorial) group field theory (GFT) approach to quantum gravity \cite{GFT1}. Two main lines of research converge in this approach: tensor models \cite{tensor} and loop quantum gravity \cite{lqg}. The first, a higher-dimensional generalisation of the matrix models formulation of 2d quantum gravity, provide the combinatorial backbone and several key mathematical tools, the second suggests the needed algebraic data that allow the definition of a Hilbert space structure and of richer, more interesting quantum geometric models for 3d and 4d quantum gravity. In these notes, as in our talk at the workshop, we illustrate briefly how this enrichment of tensor models through the input of loop quantum gravity and spin foam models works, focusing in particular on data and tools from non-commutative geometry, and what it leads to, reviewing some recent results. We thus complement the mostly combinatorial treatment of tensor models and tensorial group field theories provided by Prof. Rivasseau at the same workshop \cite{vincent3}. 

\subsection{Tensor models and why to enrich them}
Tensor models \cite{tensor} aim at a definition of quantum gravity in terms of random discrete spaces. The basic dynamical object is an array of $N_1\times ... \times N_d$ (complex) numbers $T_{i_1...i_d}$ (for simplicity, we consider $N_i = N$), where each index takes values in some finite set $i \in (1, ..., N_i)$,  which can be depicted as a (d-1)-simplex. Abusing slightly of terminology (because we do not yet assume any specific transformation property of this object), we call it a \lq tensor\rq. A choice of action functional 

$$
S(T,\bar{T}) = \bar{T} \cdot \mathcal{K} \cdot T \,+\,\lambda  \mathcal{V}( T, \bar{T}) \qquad ,
$$
where we have separated a quadratic (\lq kinetic\rq) part with kernel $\mathcal{K}$ from the higher order (\lq interaction\rq) part with kernel $\mathcal{V}$, in the simple case of a single type of interaction, weighted by coupling constant $\lambda$,
specifies then the dynamics of the system, and the quantum amplitudes obtained from the perturbative expansion (in $\lambda$) of the partition function

$$
Z \, =\, \int [dT_{i_1..i_d}][d\bar{T}_{i_1...i_d}] e^{-\, S(T,\bar{T})} \, =\, \sum_{\Gamma} \frac{\lambda^{n_{\Gamma}}}{sym(\Gamma)}\, \mathcal{A}_{\Gamma} \qquad
$$

with $n_\Gamma$ the number of interaction vertices of the Feynman diagram $\Gamma$, and $sym(\Gamma)$ the order of its eventual symmetries.
Obviously, the precise choice of action (that is, of kinematic and interaction kernels) specifies the detailed model one is studying and the structure of its Feynman diagrams $\Gamma$. The characteristic feature of tensor models (and of GFTs) is that the interaction kernels are based on a non-local pairing of tensor indices. Models with \lq\lq simplicial interactions\rq\rq like (in $d=3$) $T_{ijk} T_{klm} T_{mjn} T_{nli}$ (and complex conjugate), where repeated indices indicate complete summation over the index sets, have Feynman diagrams that can be represented by stranded diagrams (with each strand corresponding to an index of the tensor $T$ or $\bar{T}$) dual to simplicial complexes, i.e. gluings of d-simplices (corresponding to interaction kernels, e.g. a tetrahedron in the example) across shared (d-1)-simplices. One can turn $T$ into a proper tensor by assuming it transforms under the unitary (orthogonal, in the real case) group $U(N)\times ... \times U(N)$ where each component transforms a given index of the tensor $T$. A natural set of interaction kernels is then given by polynomial tensor invariants $I_\mathcal{B}(T,\bar{T})$ labelled by closed coloured graphs $\mathcal{B}$ obtained as follows: a) draw a white (resp. black) vertex for each tensor $T$ (resp. $\bar{T}$) appearing in the polynomial; b) associate $d$ different colours the $d$ indices of a tensor and draw a coloured link between a black and a white vertex corresponding to the index that is contracted between a tensor $T$ and a tensor $\bar{T}$ in the polynomial (no contraction between two $T$ or two $\bar{T}$ is allowed).  One can also associate to the coloured graph a triangulation made of a number of (d-1)-simplices equal to the order of the polynomial, glued along shared (d-2)-faces according to the pattern of links in the graph, and consider the polynomial to correspond to a d-cell bounded by such triangulation. In this case, the Feynman diagrams correspond to cellular complexes made out of glueings of such -d-cells along boundary (d-1)-simplices. In turn, any such d-cell can be triangulated by inserting a single vertex inside it and connecting this with the boundary vertices (so that the d-cell is formed by $V$ d-simplices glued around such central vertex).

The proposed definition of quantum gravity given by tensor models is therefore purely combinatorial: a sum over cellular complexes (dual to) $\Gamma$ weighted by an amplitude $\mathcal{A}_\Gamma$ that only depends on the combinatorial structure of $\Gamma$ itself. One can then endow the cellular complexes with a notion of discrete metric given by the graph distance. Then the sum over cellular complexes admits the interpretation of a sum over discrete geometries and the amplitude $\mathcal{A}_\Gamma$ can be re-interpreted (after appropriate re-interpretation of the constants $N$ and $\lambda$) as the exponential of the Regge action for the equilateral triangulation corresponding to $\Gamma$. We thus have a lattice gravity path integral analogous to the one defining the dynamical triangulations approach to quantum gravity, now generated by the field theory for $T$ and $\bar{T}$.

By now, an impressive wealth of mathematical results have accumulated for this type of models, the understanding of the combinatorics of these models, and the analytic control over their partition function is remarkable, comprising a powerful large-N expansion, some understanding of critical behaviour and double scaling limits, results on the Schwinger-Dyson equations of several models, nice universality results, and much more. For all this, we refer to the literature.

What we focus on here, instead, is the models that are obtained by enriching this already rich combinatorial models via group-theoretic data, namely \lq group field theories\rq. By this we mean models of the same tensor type but with index spaces replaced by Lie group manifolds (or subspaces thereof), as we will describe, so that one has a (complex) field $\varphi : G^{\times d} \rightarrow \mathbb{C}$ instead of the finite tensor $T$.
Why is this enrichment useful at all? One may indeed fear that it would correspond to an unnecessary complication of already complicated models, at the same time losing the beauty of a purely combinatorial definition of the fundamental structure of space and time (which is, by the way, successful in 2d as matrix models for Liouville quantum gravity show). The main advantages are the following: 

1) the definition of a proper renormalisation group flow, in turn very useful for handling the quantum dynamics of such field-theoretic models,  requires the soft breaking by the kinetic term of the invariances of the interactions (be them based on the simplicial or the tensor invariance principles); this is conveniently obtained with $\mathcal{K}$s involving differential operators which in turn require the domain of the field $T$ to be a proper differentiable manifold; (Lie) groups are an obvious choice;

2) turning the index set into a group manifold turns the tensor model into a proper quantum field theory with (a priori) an infinite number of degrees of freedom, greatly widening not only the possibilities for model building but also the set of quantum field theory tools that can be used to analyse them, and leading to richer phase structure, symmetries, etc; on the one hand, there is no proof that this widening is necessary, of course; on the other hand, it is reasonable to expect that it is, given the greater complexity of geometry and physics in dimensions higher than 2, where the purely combinatorial matrix models suffice; to face this complexity, it is nicer to have a larger variety of them at our disposal;

3) most important, the additional data allow the tensor models framework to make direct contact with other promising approaches to quantum gravity, that is loop quantum gravity and spin foam models, and thus with another large set of physical insights and mathematical results; obviously, these approaches also independently suggest that the group-theoretic data that characterise group field theories are a necessary ingredient of a fundamental description of space and time in a full quantum gravity context. Thanks to the additional data, the connection with discrete geometry and lattice gravity path integrals can also be refined, leading to further progress. And further progress can be expect via the use of tools from non-commutative geometry, which are made available by the additional data. We will now turn to a brief description of group field theories and of the above connections.

\subsection{GFTs and their connection with loop quantum gravity, spin foams and discrete gravity path integrals}
As anticipated, group field theories are field theories over a Lie group manifold $G^{\times d}$, in such a way that tensor models can be understood as a special case in which the Lie group is replaced by a finite group, e.g. $\mathbb{Z}_N$, or even its underlying finite set of N elements, neglecting the group structure entirely. We thus have the (complex) field $\varphi(g_1,...,g_d)$ as our basic variable. As such, the underlying phase space for each GFT \lq\lq quantum\rq\rq is the cotangent bundle $(\mathcal{T}^*G)^{\times d}$. In particular, for quantum gravity applications, as we will see, the most relevant groups are $G=SU(2)$ and $G= Spin(4) \simeq SU(2) \times SU(2)$ (or, in the Lorentzian signature, which will not concern us here, $G = SL(2,\mathbb{C})$). The connection with loop quantum gravity (LQG) and with lattice gravity at the kinematical level is then immediate to sketch. Canonical variables in this phase space are thus $d$ pairs of group elements and Lie algebra elements $(g_i, x_i)$, understood as classical configuration and moment variables. Given this choice of canonical variables, one can expect two basic representations for quantum states: one in terms of functions of group elements, and one in terms of conjugate Lie algebra elements. As we will discuss, the latter representation requires new mathematical tools from non-commutative geometry to be defined, due to the fact that the Lie algebra of $G$ is, in general, a non-commutative space. Another equivalent representation arises instead from the Peter-Weyl expansion of functions on the group in irreducible representations of the group itself.

The classical LQG phase space is, roughly speaking, composed of a copy of the same cotangent bundle $\mathcal{T}^*SU(2)$ for each link of all possible graphs $\gamma$. Correspondingly, quantum states of the gravitational field in loop quantum gravity \cite{lqg} are associated to all possible graphs $\gamma$ in such a way that they form a Hilbert space of functions $\mathcal{H}_\gamma = L^2(SU(2)^{E})$, with respect to the Haar measure, where $E$ is the number of links in the graph $\gamma$. Restricting in particular to d-valent graphs, one can embed each of the above Hilbert spaces $\mathcal{H}_\gamma$ into the space $\mathcal{H}_V \simeq L^2(SU(2)^{ d \times V})$, where $V$ is the number of vertices in the graph $\gamma$. In more physical language, one can consider any LQG state associated to a graph $\gamma$ as a specific \rq many-particle\rq state in which each \lq particle\rq is represented by an \lq open spin network vertex\rq, that is a graph vertex with $d$ open links attached to it, and with $d$ group elements associated to them, so that the original graph is obtained by \lq gluing\rq the open spin network vertices to one another along their links. Considering the Hilbert space for all possible number of vertices $V$ $\mathcal{H} = \oplus_V \mathcal{H}_V$, one obtains a Hilbert space large enough to contain all possible graph-based LQG states (the exact Hilbert space structure is different than in standard LQG, due to the different scalar product and the avoidance of any cylindrical consistency condition and projective limit, here). Gauge invariance conditions can be imposed in the usual way with respect to the group action at the vertices of the graphs. This last Hilbert space can be recast in 2nd quantised language as a Fock space, using the standard construction, and under the assumption of bosonic statistics for the \lq open spin network vertices\rq (other choices are of course possible). The resulting Fock space is the kinematical space of quantum states of group field theories based on this choice of group, and in absence of additional conditions on the GFT field or on its dynamics. More details can be found in \cite{fock}.

Still at the kinematical level, one realises immediately that each GFT quantum can be equally depicted as a (d-1)-simplex dual to the d-valent open spin network, similarly to tensor models, but now with its (d-2)-faces labelled by group elements $g_i$ or by Lie algebra elements $x_i$. Generic states are then going to be dual to (d-1)-simplicial complexes labelled by the same algebraic data. A first interpretation is that these are exactly the classical data of topological BF theories discretized on a simplicial complex. Indeed, when appropriate gauge invariance conditions are imposed, GFTs produce the same classical phase space of simplicial BF theories, generalised to arbitrary simplicial complexes and superpositions of them at the quantum level. These data can be understood already in terms of discrete simplicial geometry in $d=3$, where BF theory coincides with 3d gravity in the first order formalism. In this case, the Lie algebra variables correspond to discretised triad fields and the group elements to elementary parallel transports of the gauge connection (before metricity is imposed by the equations of motion). We are going to describe the GFT formulation of 3d gravity in some more detail in the following. In higher dimensions, $d=4$ in particular, BF variables can be understood in geometric terms only after appropriate geometricity conditions, called \lq simplicity constraints\rq are imposed \cite{SF}. The imposition of them in the GFT models describing BF theory in 4d, assuming this is done correctly, results in models with the right quantum geometric data to describe 4d quantum gravity. We are going to see an example of this construction in the following.

The correspondence of GFT with LQG and discrete gravity carries on to the dynamical level. Indeed, given any canonical operator equation to be imposed on LQG states and encoding the quantum dynamics of the theory, i.e. some version of the Hamiltonian constraint operator, one can write down a corresponding second quantised equation in Fock space for the GFT states; this then leads to the definition of a QFT partition function for the corresponding GFT model, assuming that one relaxes the requirement that only quantum states satisfying the quantum constraints are included in the partition function, thus working in the analogue of a grand canonical ensemble in which both energy and particle number are allowed to fluctuate (with solutions of the Hamiltonian constraint equation corresponding to \lq zero energy\rq states, and the \lq number of particles\rq being the number of graph vertices in LQG states). We have then a direct correspondence between the canonical formulation of LQG and GFTs at the dynamical level \cite{fock}. The correspondence with the covariant formulation of LQG follows suit. Given the GFT partition function so obtained, its perturbative expansion defines transition amplitudes for quantum states in terms of  a sum over cellular complexes $\Gamma$, weighted by amplitudes $\mathcal{A}_\Gamma$ which are now themselves integrals over the group (or the Lie algebra) of functions of group or Lie algebra elements (or irreps of the group) associated to the cellular complex. The latter amplitudes are, when written in terms of irreps of the group, so-called spin foam models, defining the covariant form of the dynamics of LQG. When written instead in terms of Lie algebra elements, the same amplitudes take the form of lattice gravity path integrals, that is of path integrals for simplicial gravity (like Regge calculus), written in first order form, i.e. in the classical variables of discrete (constrained) BF theory. We will see these forms of the quantum amplitudes in the following, for specific examples. 

The crucial point to note, here, is that this important connection with LQG and with discrete gravity, providing a well-defined Hilbert space as well as interesting quantum dynamical amplitudes, is possible only because of the additional group-theoretic data that enrich group field theories with respect to the simpler tensor models.

\section{Flux/Algebra representation for LQG and GFT}
The quantum geometry underlying both GFTs and LQG is manifest in the Lie algebra representation of their quantum states and amplitudes. Because in this context the Lie algebra elements are interpreted as fluxes of the gravitational triad field, this is also called {\it flux representation}. This uses some new tools from non-commutative geometry, namely a new notion of non-commutative Fourier transform.
Let us introduce these mathematical tools, before showing a few specific applications in quantum gravity. We follow the presentation in \cite{fourier}, to which we refer for further references. 

Our phase space of interest is the cotangent bundle $T^*G\cong G\times \mathfrak{g}^*$, for a Lie group $G$, the extension to several copies of this phase space being straightforward \cite{ioaristide}. With the canonical symplectic structure, we get the Poisson algebra $\mathcal{P}_G=(C^\infty(T^*G),\{\cdot,\cdot\}, \cdot)$, and for any functions $f,g \in C^\infty(T^*G)$ we obtain the brackets
$\{f,g\} \equiv \frac{\partial f}{\partial X_i} \mathcal{L}_ig - \mathcal{L}_if \frac{\partial g}{\partial X_i} + c_{ij}^{\phantom{ij}k} \frac{\partial f}{\partial X_i} \frac{\partial g}{\partial X_j} X_k\,$
where $\mathcal{L}_i$ are Lie derivatives with respect to an orthonormal basis of right-invariant vector fields, $X_i$ are Euclidean coordinates on $\mathfrak{g}^* \simeq \mathbb{R}^d$, $d:=\dim(G)$, $c_{ij}^{\phantom{ij}k}$ the structure constants, and repeated indices are summed over. We now seek to quantize a \emph{maximal subalgebra} $\mathcal{A}$ of this Poisson algebra, as an abstract operator $^*$-algebra $\mathfrak{A}$. We define a {\it quantization map} $\mathcal{Q}: \mathcal{A}\rightarrow \mathfrak{A}$ such that $\mathcal{Q}(f)=: \hat{f}$ for all $f \in \mathcal{A}_G\subset C^\infty(G)$, and $\mathcal{Q}(X_j)=:\hat{X}_j$, satisfying
$$
  [\hat{f},\hat{g}] = 0\,,\quad [\hat{X}_i,\hat{f}] = i\widehat{\mathcal{L}_i f} \in \mathfrak{A}_G\,,\quad [\hat{X}_i,\hat{X}_j] = i c_{ij}^{\phantom{ij}k}\hat{X}_k\,,\quad \forall \hat{f},\hat{g}\in \mathfrak{A}_G\,,
$$
where  $\mathcal{A}_G$ is the subalgebra of $\mathcal{A}\subset C^\infty(G\times \mathfrak{g}^*)$ of functions constant in the second argument, and $\mathfrak{A}_G := \mathcal{Q}(\mathcal{A}_G)$. In general, global coordinates cannot be defined, but can be approximated arbitrarily well by elements in $C^\infty(G)$, and we may define coordinate operators $\hat{\zeta}^i$ not necessarily in $\mathfrak{A}_G$ corresponding to a set of coordinates $\zeta^i: G \rightarrow \mathbb{R}$, satisfying  $$  [\hat{\zeta}^i,\hat{\zeta}^j] = 0 \ ,\quad [\hat{X}_i,\hat{\zeta}^j] = i\widehat{{\mathcal{L}}_i \zeta^j} \ ,\quad [\hat{X}_i,\hat{X}_j] = i c_{ij}^{\phantom{ij}k}\hat{X}_k\, . $$
Assuming $\zeta^i(e)=0$ and $\mathcal{L}_i\zeta^j(e)=\delta_i^j$, we can define $\widehat{{\mathcal{L}}_i\zeta^j} = \sum_{n=1}^{\infty} C^j_{i q_1 \cdots q_{n-1}} \hat{\zeta}^{q_1} \cdots \hat{\zeta}^{q_{n-1}}\,$,
where $C^j_{i q_1 \cdots q_{n-1}}\in\mathbb{R}$ are constant coefficients specific to the chosen coordinates. Furthermore, $U(\mathfrak{g})$ is endowed with a natural Hopf algebra structure with coproduct $\Delta_{\mathfrak{g}^*}$, counit $\epsilon_{\mathfrak{g}^*}$, and antipode $S_{\mathfrak{g}^*}$, which can be used to define a Hopf algebra structure for $\mathfrak{A}_{\mathfrak{g}^*}$, which will be crucial to define what we called the algebra representation of the quantum system. 

Given the algebra of observables, the task becomes that of finding representations of it on suitable Hilbert spaces. The simplest to define is the \emph{group representation} $\pi_G$ on $L^2(G)$ is defined as the one diagonalizing all the operators $\hat{f}\in\mathfrak{A}_G$:
$ (\pi_G(\hat{f})\psi)(g) \equiv f(g)\psi(g)\,$, for all $f\in\mathcal{A}_G$ such that $\hat{f} \equiv \mathcal{Q}(f)$. The resulting function $f\psi$ will not in general lie in $L^2(G)$ for all $\psi\in L^2(G)$, but we may restrict the domain of $\pi_G(\hat{f})$ to be the subspace of $\mathcal{A}_G$ of smooth compactly supported functions $C^\infty_c(G)$ on $G$ -- dense in $L^2(G)$ --, so that $f\psi\in C^\infty_c(G)$ for all $\psi\in C^\infty_c(G)$. Lie algebra operators $\hat{X}_i$ are represented then as
$(\pi_G(\hat{X}_i) \psi)(g) \equiv i{\mathcal{L}}_i\psi(g)\,$, with similar remarks about the domain of $\pi_G(\hat{X}_i)$. The commutation relations are then correctly reproduced. The inner product is given for $\psi,\psi'\in L^2(G)$ by
$\langle \psi,\psi' \rangle_G \equiv \int_G dg\ \overline{\psi(g)}\,\psi'(g) \,$,
where $dg$ is the right-invariant Haar measure on $G$. To prove this, one relies on the compatibility between the pointwise multiplication, used in the representation of operators, and the coproduct on the algebra. 

\subsection{Quantisation maps and algebra representation}
We now move on to define a representation in terms of functions of the classical dual space $\mathfrak{g}^*$. Obviously, the route taken to obtain the group representation cannot used because $\hat{X}_i\in\mathfrak{A}_{\mathfrak{g}^*}$ do not commute.
The strategy we adopt is then to {\it deform} the action $(\pi_{\mathfrak{g}^*}(\hat{X}_i) \tilde{\psi})(X)=X_i\tilde{\psi}(X)$, giving the needed freedom to satisfy the commutation relations. We will denote it by a star-product $\star$, and define for all $i=1,\ldots, d$
$(\pi_{\mathfrak{g}^*}(\hat{X}_i) \tilde{\psi})(X) := X_i \star \tilde{\psi}(X)\,$, in such a way that the commutator $[\hat{X}_i,\hat{X}_j] = i c_{ij}^{\phantom{ij}k}\hat{X}_k$ turns into
$ (X_i\star X_j-X_j\star X_i) \star \tilde{\psi}(X) = i\epsilon_{ijk} X_k \star \tilde{\psi}(X)\,$. In fact, we impose on the star-product the stronger condition
$(\pi_{\mathfrak{g}^*}(f(\hat{X}_i)) \tilde{\psi})(X) = f_\star(X) \star \tilde{\psi}(X)\,$, for all $f_\star\in \mathcal{A}_{\mathfrak{g}^*}\subset C^\infty(\mathfrak{g}^*)$ such that $f(\hat{X}_i) = \mathcal{Q}(f_\star) \in \mathfrak{A}_{\mathfrak{g}^*}$. This guarantees that $f_\star$ has the interpretation of the function which upon quantization gives $f(\hat{X}_i)$, and so establishes a connection between the classical phase space structure and the quantum operators. Thus,
$$(\pi_{\mathfrak{g}^*}(\mathcal{Q}(f_\star)\mathcal{Q}(f'_\star)) \tilde{\psi})(X) = (\pi_{\mathfrak{g}^*}(f(\hat{X}_i))\pi_{\mathfrak{g}^*}(f'(\hat{X}_i)) \tilde{\psi})(X) = f_\star(X) \star f'_\star(X) \star \tilde{\psi}(X) 
	= (\pi_{\mathfrak{g}^*}(\mathcal{Q}(f_\star \star f'_\star)) \tilde{\psi})(X)
$$
for all $f_\star,f'_\star\in \mathcal{A}_{\mathfrak{g}^*}$. 
Therefore, the $\star$-product and the quantization map $\mathcal{Q}$ are related by
$$
f_\star \star f'_\star = \mathcal{Q}^{-1}(\mathcal{Q}(f_\star)\mathcal{Q}(f'_\star))\,,
\label{eq:definitionstar}
$$
which is the idea of star-products defined in the context of deformation quantization and the choice of quantization map determines uniquely the $\star$-product to be used in representing the quantum algebra in terms of functions on $\mathfrak{g}^*$. Now, given a star product $\star$ and some coordinate operators $\hat{\zeta}^i$ , we define the representation of the operators $\hat{\zeta}^i$ and $\hat{X}_i$ acting on the space of smooth compactly supported functions $\tilde{\psi}\in C^\infty_c(\mathfrak{g}^*)$ on $\mathfrak{g}^*$ to be
$$
  (\pi_{\mathfrak{g}^*}(\hat{X}_i) \tilde{\psi})(X) \equiv X_i \star \tilde{\psi}(X)\,, \quad (\pi_{\mathfrak{g}^*}(\hat{\zeta}^i) \tilde{\psi})(X) \equiv -i\partial^i \tilde{\psi}(X) \,,
$$
where we denote $\partial^i := \frac{\partial}{\partial X_i}$. Now, in order to show that we have a representation of the quantum algebra, the only non-trivial part is to show that the commutator $[\hat{X}_i,\hat{\zeta}^j]$ is correctly reproduced, and this amounts to showing the compatibility between the star product and the coproduct on the observable algebra. This compatibility represent then a condition on the quantisation map to allow for an algebra representation as constructed here. The details can be found in \cite{fourier}.

\subsection{Noncommutative Fourier transform}
The next task is to find a unitary map between the two representations $\pi_G$ and $\pi_{\mathfrak{g}^*}$ of $\mathfrak{A}$. We assume that this correspondence takes the form of an integral transform, generalising the standard Fourier transform, $\mathcal{F}: L^2 (G) \rightarrow L^2_\star(\mathfrak{g}^*)$:
$$
	\tilde\psi(X):= \mathcal{F}(\psi)(X) := \int_G dg\, E(g,X)\, \psi(g) \in L^2_\star(\mathfrak{g}^*)\,,
$$
where $\psi \in L^2(G)$, and we denote by $E(g,X)$ the integral kernel of the transform. Then, the goal is to identify the defining equations for the kernel $E(g,X)$, using the fact that the intertwined functional spaces define a representation of the same quantum algebra, and applying the action of $\mathfrak{A}$ in the different representations. Its actual existence has then to be verified once an explicit choice of quantization map has been made.
We find that the kernel $E(g,X)$ si given by: 
\begin{equation}
E(g,X) = e_\star^{ik(g)\cdot X} \, = \sum_{n=0}^{\infty} \frac{1}{n!} \underbrace{ (i k(g) \cdot \{-\}) \star \cdots \star (i k(g) \cdot \{-\})}_{n\ \textrm{times}}(X),
\label{eq:Estar}
\end{equation}
with $k(g)=-i\ln(g)$ taken from any given branch of the logarithm. 
Thus, given a deformation quantization $\star$-product, this formula gives the general expression for the integral kernel $E(g,X)$. 
The same kernel should satisfy also
\begin{equation}
E(g,X) = \eta(g) e^{i\zeta(g)\cdot X}
\label{eq:Eetaexp}
\end{equation}
The prefactor $\eta(g):=E(g,0)$ depends on the quantization map $\mathcal{Q}$ chosen.
For a given $\star$-product, determining coordinates for which both forms are satisfied might be difficult and, in general, there is no guarantee that such coordinates exist. It is a requirement for the existence of the non-commutative Fourier transform. Some properties of these plane waves are:
\begin{eqnarray}
&E_g(X) = e_\star^{ik(g)\cdot X}\,=\, \eta(g)e^{i\zeta(g)\cdot X}\,, \qquad
E_e(X) = 1\,, \nonumber \\
&E_{g^{-1}}(X) = \overline{E_g(X)} = E_g(-X)\,, \qquad
E_{gh}(X) = E_g(X) \star_p E_h(X)\,, \quad
E_{g}(X+Y) = E_g(X) E_g(Y)\, \nonumber
\\
&\int_{\mathfrak{g}^*}\frac{d^dX}{(2\pi)^d}\, E_{g}(X) = \delta^d(\zeta(g)) = \delta(g) \,, \nonumber
\end{eqnarray}
where the right-hand side is the Dirac delta distribution with respect to the right-invariant Haar measure on $G$.

This completes the definition of the integral transform $\mathcal{F}$ intertwining the representations $\pi_G$ and $\pi_{\mathfrak{g}^*}$:
$$
	\tilde{\psi}(X):= \mathcal{F}(\psi)(X) = \int_G dg\, E_g(X)\,\psi(g) = \int_G dg\, e_\star^{ik(g)\cdot X}\, \psi(g) \,,
$$
where $k(g)=-i\ln(g)$ is taken in the principal branch. The $\star_p$-product is extended by linearity to the image of $\mathcal{F}$

More interesting general properties of this non-commutative Fourier transform can be found in \cite{fourier}.

Let us now give a few concrete examples of star products and plane waves following from specific choices of quantisation maps. We limit to the case of $G= SU(2)$. A generic element $k\in\su(2)$ can be written as $k=k^j\sigma_j$, $k^j\in\mathbb{R}$, where $\sigma_i$ are the Pauli matrices, while for any group element $g\in SU(2)$ we have $g=e^{ik^j\sigma_j}$. Another convenient parametrization of $\SU(2)$ can be written as
$g=p^0 \mathbb{I}+ip^i\sigma_i\,,\quad (p^0)^2+p^ip_i=1\,,\quad p^i\in\mathbb{R}\,$. These two parametrizations are thus related by the change of coordinates
$\vec{p}=\frac{\sin|\vec{k}|}{|\vec{k}|}\vec{k}\,,\quad p_0=\cos|\vec{k}|\,,\qquad k^i\in\mathbb{R}\,$, 
where $|\vec{k}|\in[0,\frac{\pi}{2}[$, or $|\vec{k}|\in[\frac{\pi}{2},\pi[$ according to $p^0\geq 0$, $p^0\leq 0$ respectively, and $g\in SU(2)$ assumes the form $
g=\cos|\vec{k}|\mathbb{I}+i\frac{\sin|\vec{k}|}{|\vec{k}|}\vec{k}\cdot\vec{\sigma}=e^{i \vec{k}\cdot \vec{\sigma}}\,$.

The so-called {\it Freidel-Livine-Majid map}, is defined, for exponentials $e^{i\vec{p}\cdot \vec{X}}$, as
$$
\mathcal{Q}_{\text{FLM}}(e^{i\vec{p}\cdot {X}}):=e^{i\frac{\sin^{-1}|\vec{p}|}{|\vec{p}|}\vec{p}\cdot \hat{{X}}}\,,
$$ and can be basically seen as the symmetrization map in conjunction with a change of parametrization.  Accordingly: 
$$
\mathcal{Q}_{\text{FLM}}^{-1}(e^{i\vec{k}\cdot\hat{{X}}})=e_\star^{i\vec{k}\cdot {X}} = e^{i\frac{\sin|\vec{k}|}{|\vec{k}|}\vec{k}\cdot{X}}\,.
$$
We may simply write $=e^{i\vec{p}(\vec{k})\cdot{X}}$, but the coordinates $\vec{p}$ only cover the upper (or lower) hemisphere $SU(2)/\mathbb{Z}_2 \cong SO(3)$, and the resulting transform is applicable only for functions on $SO(3)$.
One also finds: 
$$
e^{i\vec{p}_1\cdot{X}}\star_{\text{FLM}p} e^{i\vec{p}_2\cdot{X}} = e^{i(\vec{p}_1\oplus_p\vec{p}_2)\cdot{X}}\,\qquad \text{with}\qquad
\vec{p}_1\oplus_p\vec{p}_2=\epsilon(\vec{p}_1,\vec{p}_2)\left(\sqrt{1-|\vec{p}_2|^2}\,\vec{p}_1+\sqrt{1-|\vec{p}_1|^2}\,\vec{p}_2-\vec{p}_1\times \vec{p}_2\right)\,.
$$
The factor $\epsilon(\vec{k}_1,\vec{k}_2)=\pm 1$, introduced by the projection, is the sign of $\sqrt{1-|\vec{p}_1|^2}\sqrt{1-|\vec{p}_2|^2}-\vec{p}_1\cdot\vec{p}_2$, which is 1 if both $\vec{p}_1,\vec{p}_2$ are close to zero or one of them is infinitesimal, and $-1$ when the addition of two upper hemisphere vectors ends up in the lower hemisphere (thus projecting the result to its antipode on the upper hemisphere).

The $\star_{\text{FLM}}$-monomials read
\begin{eqnarray}
&X_i\star_{\text{FLM}} X_j = X_iX_j+i\epsilon_{ij}^{\ \ k}X_k\,,\qquad \\
&X_i\star_{\text{FLM}} X_j \star_{\text{FLM}} X_k  =X_iX_jX_k + i (\epsilon_{ijm}X_k+\epsilon_{ikm}X_j+\epsilon_{jkm}X_i)X_m+\delta_{jk}X_i-\delta_{ik}X_j+\delta_{ij}X_k\,,\quad
\ldots \nonumber
\end{eqnarray}

The non-commutative Fourier transform is then
\begin{equation}
  \tilde{\psi}(X)=\int_{\mathbb{R}^3, |\vec{p}|^2<1}\frac{d^3 {p}}{\sqrt{1-|\vec{p}|^2}}   \ e^{{i}\vec{p}\cdot {X}}\, \psi(\vec{p}) \,,
\qquad
  \psi(\vec{p}) = \sqrt{1-|\vec{p}|^2} \int_{\mathbb{R}^3} \frac{d^3 {X}}{(2\pi)^3}\ e^{{-i}\vec{p}\cdot {X}}\, \tilde{\psi}(X) \,.
\end{equation}
 
\

The {\it Duflo map} is instead given by
$$
\mathcal{D}=\mathcal{S}\circ j^{\frac{1}{2}}(\partial)\,,
$$
where $\mathcal{S}$ is the symmetric quantisation map and $j$ is the following function on $\mathfrak{g}$
$$
j(X)=\det\left(\frac{\sinh \frac{1}{2}\text{ad}_X}{\frac{1}{2}\text{ad}_X}\right)\, = \, \left(\frac{\sinh |{X}|}{|{X}|}\right)^2\,.
$$
where the last expression holds for $X\in \mathfrak{su}(2)$. When applied to exponentials, it gives
$$
\mathcal{D}(e^{i\vec{k}\cdot{X}})=\frac{\sin |\vec{k}|}{|\vec{k}|}e^{i\vec{k}\cdot\hat{{X}}}\,, 
$$
which can be inverted to
$$
\mathcal{D}^{-1}(e^{i\vec{k}\cdot\hat{{X}}})=\frac{|\vec{k}|}{\sin |{\vec{k}}|}\,e^{i\vec{k}\cdot{X}}\equiv  e_\star^{i\vec{k}\cdot{X}}\,.
$$
On monomials, we get:
\begin{eqnarray}
&X_i\star_\mathcal{D} X_j = X_iX_j+i\epsilon_{ij}^{\ \ k}X_k-\frac{1}{3}\delta_{ij}\,,\\
&X_i\star_\mathcal{D} X_j \star_\mathcal{D} X_k =X_iX_jX_k + i (\epsilon_{ijm}X_k+\epsilon_{ikm}X_j+\epsilon_{jkm}X_i)X_m+\frac{1}{3}\delta_{jk}X_i-\frac{2}{3}\delta_{ik}X_j+\frac{1}{3}\delta_{ij}X_k\,, \qquad
\ldots\nonumber
\end{eqnarray}
For the non-commutative plane wave we again have the corresponding projected star-product $\star_{\mathcal{D} p}$, which satisfies
$$
	\frac{|\vec{k}_1|}{\sin |{\vec{k}_1}|}\,e^{i\vec{k}_1\cdot{X}} \star_{\mathcal{D} p} \frac{|\vec{k}_2|}{\sin |{\vec{k}_2}|}\,e^{i\vec{k}_2\cdot{X}} = \frac{|\mathcal{B}_p(\vec{k}_1,\vec{k}_2)|}{\sin |{\mathcal{B}_p(\vec{k}_1,\vec{k}_2)}|}\,e^{i\mathcal{B}_p(\vec{k}_1,\vec{k}_2)\cdot{X}} \,,
$$

where $\mathcal{B}(k_1, k_2)$ is the Lie algebra element resulting from the BCH formula.

The explicit form of the non-commutative Fourier transform is thus
$$
  \tilde{\psi}(X)= \int_{\mathbb{R}^3,|\vec{k}|\in[0,\pi[}d^3k\, \left(\frac{\sin|\vec{k}|}{|\vec{k}|}\right)  \ e^{{i}\vec{k}\cdot {X}}\, \psi(\vec{k}) \,,
\qquad
  \psi(\vec{k}) = \int_{\mathbb{R}^3} \frac{d^3 {X}}{(2\pi)^3}\ \left(\frac{|\vec{k}|}{\sin |\vec{k}|}\right) e^{{-i}\vec{k}\cdot {X}}\, \tilde{\psi}(X) \,.
$$

We are now ready to show some applications of the algebra/flux representation to GFT, discrete gravity, spin foam models and LQG, highlighting the connections between these formalims. 

\section{GFT, spin foam models and the flux representation}
The main advantage of using the flux/algebra representation for GFT fields and amplitudes is that it brings the quantum geometry underlying them to the forefront. Indeed, as mentioned above, the algebra variables have the interpretation of discrete triad fields associated to the edges of the simplicial complexes defining the quantum states and the amplitudes, in 3d gravity, or of discrete B fields associated to triangles of simplicial complexes in 4d BF models, that become invertible for a discrete tetrad field once geometricity constraints are imposed. We will now show how this is the case.

\subsection{3d gravity: simplicial path integral, Ponzano-Regge model and semiclassical analysis}
\noindent We first consider the group field theory formulation of 3d Riemannian gravity \cite{GFT1}. The variables are fields $\varphi_{123} \!:=\!\varphi(g_1, g_2, g_3)$ on $\SO(3)^3$ satisfying the invariance:
$
\varphi_{123} = P\varphi_{123} := \int dh \, \varphi(hg_1, hg_2, hg_3) .
$
The dynamics is given by the action with simplicial (tetrahedral) interactions: 
\[
S =\! \frac{1}{2}\int [dg]^3 \,\varphi^2_{123} -  \frac{\lambda}{4!} \! \int [dg]^6 \,\varphi_{123}\varphi_{345}\varphi_{526}\varphi_{641}
\]
The Feynman diagrams $\Gamma$ are thus dual to 3d triangulations $\Delta$: the combinatorics of the field arguments in the interaction vertex is that of a tetrahedron, while the kinetic term dictates the gluing rule for tetrahedra along triangles. 

By Peter-Weyl expansion of $\varphi$ into irreps of $\SO(3)$, the field can be pictured as an open 3-valent spin network vertex or as a triangle, with the three field arguments associated to its edges $e$. In this representation, 
the interaction term can be written in terms of 6j-symbols, and the Feynman amplitude gives the Ponzano-Regge spin foam model:
$$
\mathcal{A}_\Gamma = \sum_{\{j_e\}} \prod_{e} \left( 2 j_e \,+\, 1\right) \prod_{tet} \left\{\begin{array}{ccc} j_{1} & j_{2} & j_{3} \\
	j_{4} & j_{5} & j_{6} \end{array} \right\}
$$ 
where the sum is over irreps of $\SO(3)$ labelled by integers $j_e$ associated to the edges of the triangulation, and one has a 6j-symbol for each tetrahedron in the same triangulation, function of the six representations associated to its six edges.

In group variables the same amplitudes takes the form of a lattice gauge theory imposing flatness at each edge of the simplicial complex, which is indeed the geometric content of 3d gravity and BF theory. We now make this connection even clearer by working out the algebra representation of the model \cite{ioaristide}. We use the FLM quantisation map and the associated star-product. Fourier transform and $\star$-product extend to functions of several variables like the GFT field as
$$
\hat{\varphi}_{123}\, :=\, \hat{\varphi}(x_1, x_2, x_3) =\int [dg]^3\, \varphi_{123} \, E_{g_1}(x_1) E_{g_2}(x_2) E_{g_3}(x_3) 
$$
The invariance under group action becomes the `closure constraint' for the variables $x_j$, 
$$
\widehat{P\varphi} = \widehat{C} \star \hat{\varphi}, \quad \widehat{C}(x_1, x_2, x_3) = \delta_0(x_1 \!+\!x_2\!+\!x_3), 
$$
where the non-commutative delta function has the obvious definition in terms of non-commutative plane waves; this confirms their interpretation as edge vectors (discrete triad fields) of the corresponding triangle.
The action becomes:
$$
S =\! \frac{1}{2}\int [dx]^3 \,\hat{\varphi}_{123} \star \hat{\varphi}_{123} -  
\frac{\lambda}{4!} \! \int [dx]^6 \,\hat{\varphi}_{123} \star \hat{\varphi}_{345} \star \hat{\varphi}_{526} \star \hat{\varphi}_{641}
$$
where $\star$-products relate repeated indices.
The Feynman amplitudes for closed diagrams, in this representation, read:
$$ 
\mathcal{A}_\Gamma = \int \prod_t dh_t \prod_e d^3 x_e  \, e^{i \sum_e Tr \, x_e H_e(\{ h_{t\ni e}\})}
$$
where we have used the explicit form of the FLM plane waves.
This is the simplicial path integral of first order 3d gravity (or 3d BF theory). The variables $h_t$ corresponds to the parallel transport (discrete connection) between the two tetrahedra (vertices of the Feynman diagram $\Gamma$) sharing the triangle $t$ (dual to a link of $\Gamma$); $H_e$ is the holonomy (discrete curvature) around the boundary of $f_e$ (face of $\Gamma$, dual to an edge $e$ of the simplicial complex), calculated from a chosen reference tetrahedron frame. The Lie algebra variables $x_e$, one per edge of the simplicial complex (dual to faces of $\Gamma$), play the role of discrete triad \cite{SF}. A similar result can easily be obtained for the Duflo map, leading to the same type of variables and interpretation, but to a slightly different definition of the discretized BF action, due to the different choice of coordinates on the group appearing in the plane waves (and an additional term in the measure). 

For generic open diagrams, the amplitudes are given again by the simplicial 3d gravity path integral with the appropriate boundary terms (for fixed discrete triad on the boundary).
Beside making the link with discrete geometry explicit, the flux representation is then also a useful tool for semiclassical analysis of the same amplitudes, given the fact that they take the form of standard discrete path integrals. Such analysis has been performed, for generic simplicial complexes, with metric boundary conditions, in \cite{asym}, extending previous results in the literature. We sketch now the results of this analysis, before moving on to the 4d case.
The explicit form of the Feynman amplitudes for fixed metric data on the boundary is:
\begin{eqnarray}
	\hspace{-2cm}\mathcal{A}_\Gamma(x_{ij}) &=& \int \Big[\prod_{(i,j)\in\mathcal{N}} \frac{dg_{ij}}{\kappa^3} \eta(g_{ij}^{-1})\Big] \Big[\frac{dy_{ji}}{(2\pi\hbar\kappa)^3}\Big] \Big[\prod_ldh_l\Big] \Big[\frac{d y_{e}}{(2\pi\hbar\kappa)^3}\Big] 
	\left[ \prod_{e \notin \partial\Delta} \eta(H_{e}(h_l)) \right] \left[ \prod_{\substack{(i,j)\in\mathcal{N}\\i<j}} c(g_{ij} h_{j}^{-1} K_{ji}(h_l) h_{i} g_{ji}^{-1}) \right] \nonumber \\
	&\times& \exp\left\{\frac{i}{\hbar} \left[\sum_{e \notin \partial\Delta} y_e \cdot \zeta(H_{e}(h_l)) +\!\!\! \sum_{\substack{(i,j)\in\mathcal{N}\\i<j}} y_{ji}\cdot \zeta(g_{ij} h_{j}^{-1} K_{ji}(h_l) h_{i} g_{ji}^{-1}) +\!\!\! \sum_{(i,j)\in\mathcal{N}} x_{ij}\cdot \zeta(g_{ij}^{-1}) \right]\right\} \qquad
\end{eqnarray}
where: the set of ordered pairs of labels associated to neighboring boundary triangles is $\mathcal{N}$; $h_i$ is the group element associated to the dual half-link going from the boundary triangle $i$ to the center of the bulk tetrahedron with triangle $i$ on its boundary; $K_{ij}(h_l)$ is the holonomy along the bulk dual links from the center of the tetrahedron with triangle $j$ to the center of the tetrahedron with triangle $i$; $x_{ij}$ is the discrete triad (edge vector) shared by the triangles $i,j$ on the boundary, as seen from the frame of reference of the triangle $j$; $\kappa=8\pi G$ is the non-commutativity parameter entering star products and plane waves (so the abelian limit of the phase space corresponds to the no-gravity limit $G\rightarrow 0$), the $y_e$ are discrete triad variables associated to edges in the bulk of the triangulation, while $y_{ij}$ are auxiliary metric variables associated to boundary edges (they end up being identified with the $x_{ij}$ by the dynamics, as it can verified immediately by performing some of the group integrations); $\eta(g)$ are real functions on the group depending on the choice of quantisation map (one has $\eta(g)=1$ for the FLM map and $\eta(g) = \frac{|k(g)|}{\sin|k(g)|}$ for the Duflo map, as we have seen).
The amplitudes are directly suited for studying the semiclassical limit through a saddle point analysis of the discrete gravity action:
$$
\mathcal{S} := \sum_{e \notin \partial\Delta} y_e \cdot \zeta(H_{e}(h_l))\ + \sum_{\substack{(i,j)\in\mathcal{N}\\i<j}} y_{ji}\cdot \zeta(g_{ij} h_{j}^{-1} K_{ji}(h_l) h_{i} g_{ji}^{-1})\ + \sum_{(i,j)\in\mathcal{N}} x_{ij}\cdot \zeta(g_{ij}^{-1}) 
$$
in first order variables $y_e$ and $h_l$, plus boundary data $x_{ij}$, $y_{ij}$ and $g_{ij}$ (only the first are fixed at the boundary, the others being subject to variations). We expect the classical limit to be governed by the equations of motion of this action:  geometricity constraints imposing flatness of holonomies around dual faces and closure of edge vectors for all triangles (up to parallel transport).
This is exactly what happens. Standard variations of this action give the equations:
$$
		\zeta(H_{e}(h_l)) = 0 \Leftrightarrow H_{e}(h_l) = \mathbb{I}
$$	
for all $e\notin\partial\Delta$, i.e., the flatness of the connection around the dual faces in the bulk. 
$$
\zeta(g_{ij} h_{j}^{-1} K_{ji}(h_l) h_{i} g_{ji}^{-1}) = 0 \Leftrightarrow g_{ij} h_{j}^{-1} K_{ji}(h_l) h_{i} g_{ji}^{-1} = \mathbb{I}
$$
	for all $(i,j)\in\mathcal{N}, i<j$, i.e., the triviality of the connection around the dual faces to $e\in\partial\Delta$. 
$$	
	\sum_{\substack{e \in \Delta\\e^*\ni f^*}} \epsilon_{fe}(\textrm{Ad}_{G_{fe}} y_e) = 0  \qquad \sum_{\substack{f_j\in\partial\Delta\\(i,j)\in\mathcal{N}}} \epsilon_{ji} (\textrm{Ad}_{g_{ji}}^{-1}y_{ji}) = 0 
$$
where $\textrm{Ad}_{G_{fe}}$ implements the parallel transport from the frame of $y_e$ to the frame of $f$, and $\epsilon_{fe}=\pm 1$ accounts for the orientation of $h_l$ with respect to the holonomy $H_{e^*}(h_l)$ and thus the relative orientations of the edge vectors, and  $\textrm{Ad}_{g_{ji}}^{-1}$ parallel transports the edge vectors $y_{ji}$ to the frame of the boundary triangle $f_i$, and $\epsilon_{ji}= \pm 1$ again accounts for the relative orientation. These impose the closure constraint for the three edge vectors of each bulk triangle $f \notin \partial\Delta$ in the frame of $f$, and the closure of the boundary integration variables $y_{ji}$. These same conditions give the metric compatibility of the discrete connection, which in turn, if substituted back in the classical action, before considering the other saddle point equations, turn the discrete 1st order action into the 2nd order Regge action for the triangulation $\Delta$.

\begin{equation}\label{eq:defXclosure}
	\sum_{\substack{f_j\in\partial\Delta\\(i,j)\in\mathcal{N}}} D^{\zeta}(g_{ij}) x_{ij} = 0\ \forall i  \qquad \qquad \textrm{Ad}_{g_{ij}} (D^\zeta(g_{ij})x_{ij}) = -\textrm{Ad}_{g_{ji}} (D^\zeta(g_{ji})x_{ji}) 
\end{equation}
where we denote $(D^{\zeta}(g))_{kl} := \tilde{\mathcal{L}}_k\zeta_l(g)$; these are a \emph{deformed} closure constraint for the boundary edge variables $x_{ij}$, and a \emph{deformed} identification, up to a parallel transport, of the boundary edge variables $x_{ij}$ and $x_{ji}$.
The amplitudes become
\begin{eqnarray}
	&\tilde{\mathcal{A}}_\Gamma(x_{ij}) \propto \int \Big[\prod_{(i,j)\in\mathcal{N}} \frac{dg_{ij}}{\kappa^3} \eta(g_{ij}^{-1})\Big] \Big[\prod_{v\in\partial\Delta} \delta(H_v(g_{ij})) \Big] \left[ \prod_{f_i \in \partial\Delta} \delta_\star\Big(\!\!\!\sum_{\substack{f_j\in\partial\Delta\\(i,j)\in\mathcal{N}}} \!\!\! D^{\zeta}(g_{ij}) x_{ij}\Big) \right] \nonumber \\
	& \star \left[ \prod_{\substack{(i,j)\in\mathcal{N}\\i<j}} \delta_\star\Big(\textrm{Ad}_{g_{ij}} (D^\zeta(g_{ij})x_{ij}) + \textrm{Ad}_{g_{ji}} (D^\zeta(g_{ji})x_{ji})\Big) \right] \star\ \exp\left\{\frac{i}{\hbar} \sum_{(i,j)\in\mathcal{N}} x_{ij}\cdot \zeta(g_{ij}^{-1}) \right\}  \big(1 + \mathcal{O}(\hbar)\big)  \qquad\label{eq:0thPRamp}
\end{eqnarray}
where the delta functions impose the constraints on boundary data discussed above. In particular, $H_v(g_{ij})$ are the holonomies around the boundary vertices $v\in\partial\Delta$, whose triviality follows from the triviality of the bulk holonomies. Notice that one must write the integrand in terms of $\star$-products and $\star$-delta functions in order for the constraints to be correctly imposed, since the amplitude acts on wave functions through $\star$-multiplication. The non-commutative nature of the variables actually has to be taken into account in a more subtle way. In fact, it turns out \cite{asym}, that one needs to consider the deformation of phase space structure and take a {\it non-commutative variation} $\delta_\star S$ of the action $S$ in the amplitude via $e_\star^{i\delta_\star S + \mathcal{O}(\delta^2)} \equiv e_\star^{iS^\delta} \star e_\star^{-iS}$, where the $\star$-product acts on the fixed boundary variables $x_{ij}$, $\mathcal{O}(\delta^2)$ refers to terms higher than first order in the variations, and $S^\delta$ is the varied action. It is easy to see that the non-commutative variation so defined undeforms the identification of $x_{ij}$ and $x_{ji}$ (up to parallel transport), simply because we have
\begin{equation}
	E(e^{i\epsilon Z}g,x) \star E(g^{-1},x) = E(e^{i\epsilon Z},x) \star \underbrace{E(g,x) \star E(g^{-1},x)}_{=1} = e^{\frac{i}{\hbar\kappa}\epsilon (Z\cdot x) + \mathcal{O}(\epsilon^2)}
\end{equation}
for any $Z\in \su(2)$ and $\epsilon \in \mathbb{R}$ implementing the variation of $g$. We obtain the geometric identification  $\textrm{Ad}_{g_{ij}}^{-1} y_{ji} = \pm x_{ij}$, and the undeformed closure condition for boundary $x_{ij}$. These undeformed relations replace the deformed ones in the above semiclassical approximation of the amplitudes. We recover exactly the geometric relations for the boundary variables, regardless of the choice of a quantization map, and the standard equations of motion for 3d gravity in the bulk.
Once more, the algebra representation of the theory allows for a straightforward, geometrically transparent analysis.

\subsection{A model for Holst-Plebanski 4d gravity}
We now turn to 4d quantum gravity and describe the application of the non-commutative geometry tools introduced above to the construction of interesting GFT models of it. We will see another instance of the general duality between spin foam models and simplicial gravity path integrals, and of the way the algebra/flux representation brings the quantum geometry of such models to the forefront. More details can be found in \cite{GFTimmirzi}.

Classical continuum 4d gravity  can be expressed as a constrained $BF$ theory \cite{GFT1, SF} 
\begin{equation} \label{pleb}
S(\omega,B, \lambda)=\int_{\mathcal{M}} Tr \, B \wedge F (\omega) + \lambda \, \mathcal{C}(B), 
\end{equation}
for $\so(4)$ valued 1-form connection $\omega$ and 2-form field $B$ (also Lie algebra valued),  where $\mathcal{C}(B)$ are (so-called) simplicity constraints  and $\lambda$ is 
some  Lagrange multiplier. The constraints force $B$ to be 
a function of a tetrad 1-form field 
$B\!=\!\ast (e\wedge e)$, turning  BF to the Palatini action for gravity in the first order formalism. 
The Immirzi parameter $\gamma$, which plays a crucial role in LQG, is introduced by changing variables $B \!\to\! B + \frac{1}{\gamma}\!\ast\! B$ in the BF term. 
The constraints lead then to the Holst action,  classically equivalent to Palatini gravity and the classical starting point of LQG.

The general strategy in group field theory and spin foam models \cite{GFT1, SF} has been, therefore, to start from a formulation of simplicial 4d BF theory and impose a discrete counterpart of the simplicity constraints to obtain a model for 4d gravity.
The starting point is therefore the straightforward 4d generalisation of the GFT model for 3d gravity described above:
$$
S = \frac{1}{2} \int [dg_i]^4 \varphi^2_{1234}  + \frac{\lambda}{5!}  \int [dg_i]^{10}  
\varphi_{1234} \,\varphi_{4567} \,\varphi_{7389}\, \varphi_{962\,10} \,\varphi_{10\,851} 
$$
where $\varphi_{1234}=\varphi(g_1, \cdots, g_4) = \varphi(h g_1, \cdots, h g_4)$ and $g_i, h \in \SO(4)$. The perturbative expansion in $\lambda$ of the partition function generates a sum over (closed) Feynman diagrams which are cellular complexes dual to simplicial 4d complexes, to which the theory assigns amplitudes which, once more, can be written down equivalently as a BF lattice gauge theory (in the group representation), as simplicial path integrals for discrete BF (in the algebra representation) and as a spin foam model (in the spin representation). The algebra representation is obtained by extending the non-commutative $\SO(3)$ Fourier transform 
to $[\SO(3)\times \SO(3)]^4$ (in the following, we use the FLM quantisation map), and leads to a function of four $\so(4)$ elements $\varphi(x_1,...,x_4)$ interpreted as the discrete B fields of BF theory, associated to the four triangles belonging to the tetrahedron corresponding to the GFT field. The gauge invariance condition becomes again the closure $x_1+..+x_4=0$. An entirely equivalent model is obtained by adding an additional variable $k\in S^3\simeq \SU(2)$, interpreted as unit vectors normal to the tetrahedron corresponding to the GFT field, relaxing the gauge invariance condition to the covariance: $\varphi_k(g_1, \cdots g_4)= \varphi_{h \triangleright k}(h g_1, \cdots h g_4)$, $\forall h \in \SO(4)$ (using the selfdual/anti-selfdual decomposition of $\SO(4)$, $h \triangleright k:=h^{+} k (h^{-})^{- 1}$ is the normal rotated by $h$) and defining the action: 
$$
S[\varphi]  = \frac{1}{2}\int  [dg_i]^4 dk \, \varphi^2_{k1234}  + \frac{\lambda}{5!} \! \int [dg_i]^{10} [dk_i]^5 \, 
\varphi_{k_11234} \varphi_{k_24567} \varphi_{k_37389} \varphi_{k_4962\,10} \varphi_{k_510\,851} \; .
$$
The discrete counterpart of the simplicity constraints is, classically \cite{GFTimmirzi, SF, GFT1}:
\begin{equation} \label{simp}
\forall j\in \{1...4\}, \quad \exists k \in \SU(2), \quad  kx^-_j k^{- 1} + \beta x^+_j = 0
\end{equation}
where
$\beta\!=\!\frac{\gamma-1}{\gamma+1}$. A geometrically clear way of imposing these constraints in the model is to use non-commutative delta functions, thus effectively
constraining the measure on the bivectors. 
We thus introduce the function $S_k^\beta(x) := \delta_{- kx^- k^{-1}}(\beta x^+) = \int_{\SU(2)} \!\!du \, e^{i \mbox{\footnotesize tr} [k^{-1} u k x^-]} e^{i \beta  \mbox{\footnotesize tr} [u x^+]}= \int_{\SU(2)} \!\!\extd u \, E_{\mathbf{u}^k_{\beta}}(x)$
where: $u_\beta\!=\! e^{\theta_{\beta} n^j_{\beta} \tau_j}$, with $\theta_\beta$ and $\vec{n}_\beta$ such that $\sin\theta_\beta \!=\! |\beta| \sin \theta, \,\, \mbox{\footnotesize sign}(\cos\theta_\beta) = \mbox{\footnotesize sign}(\cos\theta); \quad \vec{n}_{\beta} = \mbox{\footnotesize sign}(\beta) \vec{n}$,  
 we introduced $\mathbf{u}^k_{\beta} \!=\! (k^{-1} u k, u_\beta) \!\in\! \SU(2) \!\times\! \SU(2)$, and $\delta_{- a}(b)\!:=\!\delta(a+b)$.  
The action of $\widehat{S}$ is well-defined on gauge invariant fields, as it commutes with the gauge transformations
$
\widehat{S}^\beta\triangleright [E_h \cdots E_h \star \hat{\varphi}_{h^{-1}\triangleright k}] = E_h \cdots E_h \star (\widehat{S}^\beta\triangleright\hat{\varphi})_{h^{-1} \triangleright k}
$
These relations ensure that rotating a  bi-vector which is simple with respect to a normal $k$ gives  a bi-vector which is simple with respect to the rotated normal, as geometrically required.

A geometrical GFT model will be defined by constraining the field $\hat{\varphi}_k(x_j)$, by acting on it  by $\star$-multiplication by the product
$S_k^\beta(x_1)...S_k^\beta(x_4)$ of four simplicity functions:
\begin{equation} \label{simplicity}
(\widehat{S}^\beta \triangleright \hat{\varphi})_k(x_1, \cdots, x_4) = \prod_{j=1}^4 S_k^\beta(x_j) \star \hat{\varphi}_k (x_1,\dots x_4) 
\end{equation}

For generic values of $\beta\!=\!\frac{\gamma-1}{\gamma+1}$,  the operator $\widehat{S}^\beta$ is not a projector. 
Depending on whether it is inserted in the propagator, in the vertex, or in both, in a single or in multiple copies, 
one gets slightly different Feynman amplitudes. We choose to constrain the field in the interaction of the extended BF model. Defining 
$\widehat{\Psi}^\beta\!:=\! \int dk \widehat{S}^\beta\triangleright \hat{\varphi}_k$, we consider: 
\begin{equation} \label{actionbeta}
S \!\!=\!\!\frac{1}{2} \int [d^6 x_i]^4 \,dk  \, \widehat{\varphi}_{k 1 2 3 4} \star \widehat{\varphi}_{k 1 2 3 4} 
\;  + \frac{\lambda}{5!}  \int [d^6 x_i]^{10}   \,
\widehat{\Psi}_{1 2 3 4} \star \widehat{\Psi}_{4 5 6 7}\star  \widehat{\Psi}_{7 3 8 9}\star  \widehat{\Psi}_{9 6 2\,10} \star \widehat{\Psi}_{10\,8 5 1} 
\end{equation}
where  the star product pairs repeated indices. We can now compute the Feynman amplitudes of this model.
One sees that, for each triangle $t$, the constraints impose the linear simplicity condition of $x_t$ with respect to the normals of all the tetrahedra $\{\tau_j\}_{j=0...N_t}$ sharing $t$.  
The square is present because both 4-simplices $\sigma$ sharing the tetrahedron $\tau_j$ contribute a factor $S^{\beta}  _{h_{0j} \triangleright k_j}(x_t)$. 
For closed graphs, the normals $k_{\tau} \!\in\! \SU(2)$ drop from the amplitude and we obtain: 
\begin{equation} \label{explPI}
\mathcal{A}_\Gamma(\beta) = \int  [dh_{\tau\sigma}][d^6 x_t] \left[\prod_t \bigstar_{j=0}^{N_t} 
\delta^{\star 2}_{- \bar{h}_{0j} x^-  \bar{h}_{0j} ^{-1}}(\beta x^+)\right]  \star e^{i \sum_t  Tr \, x_t H_t}
\end{equation}
where we wrote $f^{\star 2}$ for the squared function $f\star f$, and $\bar{h}_{0j}\!=\!h_{0j}^+ (h^-_{0j})^{- 1}$. The simplicity operators for the same triangle, appearing in different frames corresponding to different tetrahedra sharing it, can be split into a single constraint expressed in a given frame, and a modified measure term for the discrete connection, encoding the conditions on the connection that have to be satisfied to ensure the correct parallel transport of the simplicity constraint across simplicial frames.

The  Feynman amplitudes of this theory thus take the form of simplicial path integrals for a constrained BF theory of Holst-Plebanski type with Immirzi parameter $\gamma$, with linear simplicity constraints. As noted, the specific form of the modification of the BF measure, encoding the simplicity constraints, depends on how exactly one has inserted the simplicity operator in the GFT action. The same amplitudes can be written down in pure gauge theory form (in the group representation) and in pure spin foam form (expanding in irreducible representations). In the latter form can be compared more easily with other spin foam models for 4d gravity proposed in the literature \cite{SF}. We refer to \cite{GFTimmirzi} for more details. 

\section{Conclusions}
We have presented a brief review of the role of group-theoretic data in group field theories, focusing on the non-commutative flux representation of gravity models. While we believe that this could be of interest to mathematicians for its own sake, we have shown how useful it is from a quantum gravity point of view. It makes manifest the quantum geometric content of GFTs, and realises an explicit duality between spin foam models and simplicial gravity path integrals at the level of GFT amplitudes, providing an rich, promising framework for quantum gravity.

\begin{acknowledgement}
We thank the organisers of the workshop on Noncommutative Field Theory and Gravity in Corfu, for their hospitality and for a really enjoyable and interesting event.
\end{acknowledgement}

\end{document}